\documentclass{emulateapj}
\usepackage{times}
\usepackage{graphicx}

\slugcomment{Submitted to ApJ }

\shorttitle{B-ducted heating of Black Widows}
\shortauthors{Sanchez \& Romani }

\begin{document}

\title{B-ducted Heating of Black Widow Companions}

\author{Nicolas Sanchez \& Roger W. Romani \altaffilmark{1}}
\altaffiltext{1}{Department of Physics, Stanford University, Stanford, CA 94305-4060,
 USA; rwr@astro.stanford.edu}

\begin{abstract}

The companions of evaporating binary pulsars (black widows and related systems)
show optical emission suggesting strong heating. In a number of cases large observed temperatures
and asymmetries are inconsistent with direct radiative heating for the observed pulsar spindown power
and expected distance.  Here we describe a heating model in which the 
pulsar wind sets up an intrabinary shock (IBS) against the companion 
wind and magnetic field, and a portion of the shock particles duct along this field 
to the companion magnetic poles. We show that a variety of
heating patterns, and improved fits to the observed light curves, can be obtained at 
expected pulsar distances and luminosities, at the expense of a handful of model parameters.
We test this `IBS-B' model against three well observed binaries and comment on the implications
for system masses.
\end{abstract}


\section{Introduction}

	When PSR J1959+2048 was found to have a strongly heated low mass companion,
visible in the optical, it was realized that this heating was an important probe of
the millisecond pulsar (MSP) wind \citep{de88,cvr95} and that study of the companion dynamics
gave a tool to probe the MSP mass \citep{arc92,vKBK11}. The discovery that many {\it Fermi} LAT
sources were such evaporating pulsars alerted us to the fact that this wind-driving phase
was not a rare phenomenon, but a common important phase of close binary pulsar evolution.
It has also provided a range of examples from the classical black widows (BW) like PSR J1959+2048
with $0.02-0.04M_\odot$ companions, to the redbacks (RB) with $0.2-0.4M_\odot$ companions,
to the extreme `Tiddarens' (Ti) with $\le 0.01 M_\odot$ He companions \citep{rgfz16}. Several
of these systems are near enough and bright enough to allow detailed optical photometry
and spectroscopy. The results have been puzzling: the simple models that imply direct radiative
pulsar heating of the companions' day sides, while sufficient for the early crude measurements
of simple optical modulation, often do not suffice to explain the colors, asymmetries 
\citep{stap01,sh14} and heating efficiency revealed by higher quality observations
\citep{rfc15}.

These binaries are expected to 
sport an IntraBinary Shock (IBS) between the baryonic companion wind and the relativistic
pulsar wind. This shock certainly reprocesses the relativistic particle/$B$ wind of the pulsar
and Romani \& Sanchez (2016, hereafter RS16) showed how radiation from the IBS could create
the characteristic peak structures seen in RB and BW X-ray light curves \citep{robet14} and, in 
some cases, asymmetric surface heating.

	However extreme asymmetries and high required efficiency
were a challenge for this model, and it was suspected that
direct particle heating of the companion surface might play an important role. Here we describe
such heating, where a portion of the IBS shock particles cross the contact discontinuity
and are `ducted' to the companion surface along the
open pole of a companion-anchored magnetic field. Such fields are expected since companion
tidal locking ensures rapid rotation in these short-period binaries, while the presence of
emission from the night side of the orbit suggests sub-photospheric convective heat transport;
rotation and convection being the traditional ingredients for a stellar magnetic dynamo.
Theoretical \citep[e.g.][]{a92} and observational \citep[e.g.][]{rfc15,vsa16} evidence has been presented
for such companion fields in black widows and related binaries. The idea that these fields
could enhance BW heating was first described by \citet{eich92} for PSR J1959+2048, where it was
suggested that Alfv\'en wave propagation in the companion magnetosphere might, in analogy
with the Sun-Earth system, increase the heating power by as much as $100\times$
over direct radiation.

	We present a simple model for such `IBS-B' heating assuming a dominant dipole field
and implement it in a binary light curve fitting code. We show that this model explains
a number of peculiarities in high-quality BW light curves. Several example fits are shown.
In some cases extension beyond the simple dipole estimate may be required. However, the fits
are good enough so that conclusions drawn from direct heating fits should be viewed with
caution. Precision neutron star mass measurements will therefore require high quality 
companion observations and robust heating models.

\section{Basic Model}

In direct heating the pulsar spindown power radiatively heats the facing side of the companion,
raising the characteristic temperature from the unheated (`Night' side) $T_N$ to
$$
T_D^4=\eta {\dot E}/4\pi a^2 \sigma_{SB} +T_N^4
\eqno (1)
$$
with $a=x_1 (1+q)/{\rm sin}\,i$ the orbital separation,
$x_1$ the projected semi-major axis of the pulsar orbit, ${\dot E}=I\Omega{\dot \Omega}$ the
pulsar spindown power for moment of inertia $I$ (spin angular frequency and derivative
$\Omega$ and ${\dot \Omega}$), Stefan-Boltzmann constant $\sigma_{SB}$ and $\eta$ 
a heating efficiency. Effectively $\eta=1$ corresponds to isotropic pulsar emission 
of the full spindown power, with the impingent radiation fully absorbed by the companion.
This model has been implemented in several light curve modeling codes
eg. the ELC code \citep{oh00} and its descendant ICARUS \citep{bet13}. Such direct heating
certainly describes the effect of the pulsar photon irradiation.  Yet the observed
radiative flux, peaking in GeV $\gamma$-rays, represents only a fraction of the pulsar
spin-down power, with a heuristic scaling $L_\gamma \approx ({\dot E} \cdot 10^{33}{\rm erg/s})^{1/2}$.
The bulk of this power is instead carried in a $e^+/e^-/B$ pulsar wind.

	If the MSP wind terminates in an IBS, this may contribute to
the surface heating (RS16). In that paper it was assumed that the beamed
emission from the oblique relativistic shock was the primary heating source. 
Such radiation does indeed produced the double-peaked X-ray light curves of BW, RB
and related binaries, with the X-ray peaks caustics from synchrotron emission beamed 
along the IBS surface. Two parameters control the IBS shape: the ratio between the
companion and MSP wind momentum fluxes $\beta$, and the speed of the companion wind relative
to the its orbital velocity $v_w=f_v v_{orb}$. Here $\beta = {\dot m}_W v_w c /{\dot E}$ sets 
the scale of the IBS shock with a characteristic standoff
$$
r_0= {{\beta^{1/2}} \over {(1+\beta^{1/2})}} a;
\eqno (2)
$$
with $\beta<1$ IBS wrapping around the companion and $\beta>1$ shocks surrounding the pulsar.
When $f_v \sim 1$ or smaller, the IBS is swept back in an Archemidean spiral and the X-ray 
light curves and surface heating are asymmetric.

Although simple estimates indicate that the post-shock particles can cool
on the IBS flow time scale and the shocked particles have a modest bulk $\Gamma$ 
\citep{boget12} so that some of this cooling radiation can reach the companion, the general effect of
an IBS shock is to deflect pulsar power {\it away} from the companion. In addition
the observed IBS radiation, while a substantial portion of the the X-ray band flux, is only 
$L_X \sim 10^{-3}-10^{-4} {\dot E}$ \citep{apg15}. Since IBS X-rays are emitted closer to
the companion than the pulsar gamma-rays, they can be more effective at companion heating by
$\sim \beta ^{-1}$. However the observed ratios are typically $f_\gamma/f_X \sim 300$ (Table 1),
so it seems difficult for IBS emission to dominate companion heating, unless
$\beta$ is very small, or the IBS emission is primarily outside the X-ray band or is 
beamed more tightly to the companion than the pulsar emission. So while IBS radiation is doubtless
useful for a detailed treatment of companion heating (and may contribute needed asymmetry),
it is unlikely to dominate. The problem is particularly severe for systems whose observed $T_D$ indicates
a higher heating power than that emitted by the pulsar into the solid angle subtended by the companion.

	The solution appears to be more direct utilization of the pulsar wind. We will
assume here that as the pulsar wind impacts (either directly or at a wind-induced
IBS) a companion magnetic field a fraction of the particles thread the companion
field lines and are ducted to the surface. Since the field and flow structure in
the IBS are poorly understood, we do not attempt here to follow the details of how these
particles load onto the companion magnetosphere. As noted in RS16, companion
fields of plausible strength can be dynamically important at the IBS. In 
the small $\beta$ limit a companion with radius $r_\ast$ will have a surface dipole field 
$B_c$ dominating at the IBS for
$B_c > (2 \beta {\dot E}/c) ^{1/2} \beta a^2/r_\ast^3$. With typical $a\sim 10^{11}$cm
and $r_\ast \sim 0.1 R_\odot$ this is $B_c \sim 24 \beta^{3/2}$kG, which seems plausible
for the large scale dipole of such rapidly rotating stars. However, given the evidence for
a strong evaporative wind, we envision that this
contributes significantly to the IBS-forming momentum flux,
so the dipole field may be somewhat lower, and the wind controls the IBS shape. 
For larger $B_c$ the strength and orientation of the companion field will be important
to the details of the IBS shape, although the basic energetics should be close to
that of the geometry considered here. Note that even a weaker field can still
duct the energetic $e^+/e^-$ of the shocked pulsar wind as long as the gyroradius
is $< r_0$. For $r_0 \sim \beta^{1/2} a$ the condition on particle energy for ducting
is $\gamma < r_0B_c(r_\ast/r_0)^3/1.7\times 10^3$cm $\sim 2 \times 10^4 B_c/\beta$, with
$B_c$ in Gauss. At the IBS-controlling $B_c$ above this is $\gamma < 4 \times 10^8 \beta^{1/2}$,
so if the post-shock energy flux is dominated by, say $\gamma < 10^6$, then fields
$\sim 100\times$ weaker can still effectively duct. 

	A more immediate issue is whether
these ducted particles can deposit the bulk of their energy at the field foot points.
In general the fields are too weak for efficient synchrotron cooling on the inflow
timescale with a mirror to synchrotron ratio
$\zeta \sim 10^{15}\beta^{5/2}B_c/(a/10^{11}{\rm cm})$ \citep{ho86}.
Thus even with $B_c$ important for the IBS structure this is 
$\zeta \sim 10^6 \beta^{5/2} (B_c/24{\rm kG})^{-2} (a/10^{11}{\rm cm})^{-1}$, so in an empty magnetosphere the 
$e^+/e^-$ will mirror above the pole unless $\beta$ is very small. Mirroring particles
can radiate energy transverse to the companion surface and may be lost from
the magnetosphere.  However, if the postshock distribution is anisotropic, the fraction 
entering with small perpendicular momentum and precipitating to the pole may not be small. 
Also mirrored particles may be captured and mirror between poles until
they precipitate. Finally other damping may be important, especially since the 
wind should flow out along the heated companion pole. 
A simple estimate for the column density traversed by 
the precipitating particles is $\beta {\dot E}/(\theta_c^2 c v_w^2 r m_P)$
$\sim 10^{22} \beta/[\theta_c^2 (v/500 {\rm km/s})^2 (r/10^{10} {\rm cm})] {\rm cm^{-2}}$. 
With a modest $\beta$ 
the (forward beamed) energy losses will slow a significant fraction of the precipitating
particles, enhancing capture, especially if mirroring makes multiple reflections
to the forward pole with $\theta_c < 1$. In any event, our geometrical model computes the
effect of precipitating particle energy thermalized in the magnetic cap. If this
is not a large fraction of the spindown power incident on the open zone above these caps,
the particle heating effect appears plausible. More detailed sums will be needed to
establish the true efficacy.

\begin{figure}[t!!]
\vskip 10.1truecm
\includegraphics{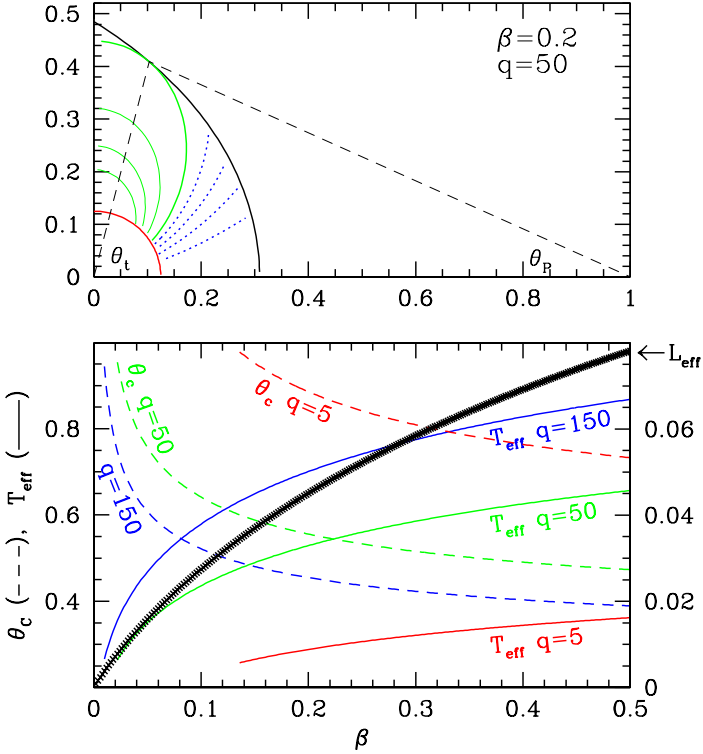}
\begin{center}
\caption{\label{IBS-B} 
Above: Intra-Binary Shock enclosing an aligned dipole field. Here BW-like
parameters $\beta$ and $q$ are assumed. The PSR emission within $\theta_P$
impacts the open zone above the companion cap subtending $\theta_t = 1.324$.
Below: The fraction of pulsar spin down power impingent on the forward cap
for a Dyson IBS approximation, isotropic MSP wind and aligned companion dipole
as a function of $\beta$ (thick black line, right scale). 
Colored curves give the $\beta$-dependence of the
cap footpoint angle $\theta_c$ in radians (dashed lines, left scale) and the 
mean cap temperature (solid lines, left scale units see Eq. 8)
for three different mass ratios: $q=5$ (RB-like, red), $q=50$ (BW-like, 
green) and $q=150$ (Ti-like, blue). The curves stop at small $\beta$, when the
IBS standoff touches the Roche lobe-filling star.
}
\end{center}
\vskip -0.5truecm
\end{figure}

To obtain the basic scaling in this model we start with a simplified symmetric bow shock geometry
$$
r_D(\theta) = r_0 \theta/{\rm sin}\theta
\eqno (3)
$$
\citep{dy75}, and describe the heating on the pulsar-facing side of the companion which
has a spherically symmetric pulsar wind and a dipole field
$$
r_B(\theta) = r_\ast {\rm sin}^2\theta/{\rm sin}^2\theta_C
\eqno (4)
$$
whose axis is aligned with the line of centers and which is anchored in a Roche-lobe
filling companion of size $r_\ast = 0.46 q^{-1/3} a$, with $q=m_{P}/M_{c}$ the mass ratio
and $a$ the orbital separation. The cap half angle $\theta_c$ describes the footpoints 
of the open field lines. This open zone is determined by the intersection of the 
field lines and the IBS; these are tangent when
$$
\theta_t = ({\rm tan}\,\theta_t)/3, \qquad \qquad i.e.\quad \theta_t=1.324
\eqno (5)
$$
which implicitly determines the portion of the pulsar wind striking the open zone. If
a fraction $f_{\dot E}$ of the power into this zone reaches the cap we have a heating luminosity
$$
L_h = f_{\dot E} {\dot E} (1-{\rm cos}\theta_P)/2,
\eqno (6)
$$
where from the MSP the open zone subtends ${\rm cos}\theta_P= ({\rm sin}\theta_t -r_0\theta_t{\rm cos}\theta_t)/
(r_0^2\theta_t^2+{\rm sin}^2\theta_t-r_0 \theta_t {\rm sin}2\theta_t)^{1/2}$
(Figure 1). The area of the heated surface is
$$
A_c=2\pi r_\ast^2 (1-{\rm cos}\theta_C)
\eqno (7)
$$ 
with $1-{\rm cos}\theta_C=1-[1-(r_\ast {\rm sin}^3 \theta_t)/(r_0\theta_t)]^{1/2}$.
This gives the heating power, cap size and the average cap temperature 
$T_{eff}= (L_h/\sigma_B A_c)^{1/4}$. These quantities are shown for this simple
model in Figure 1, with an assumed $f_{\dot E} =0.01$ and $T_{eff}$ in units of 
$$
(L_h/\sigma_B a^2)^{1/4} = 13,800\,K\, {\dot E}_{34}^{1/4} a_\odot^{-1/2}
\eqno (8)
$$
for characteristic MSP power ${\dot E} = {\dot E}_{34} 10^{34} {\rm erg\,s^{-1}}$,
and orbital separation $a_\odot R_\odot$.

	Note that the heating power increases and the cap size decreases with $\beta$,
leading to an increase in $T_{eff}$ for strong companion winds. Indeed, it is
believed that the heating drives the companion wind and so magnetic-induced winds may
in fact have positive feedback, with stronger heating leading to larger outflow and
larger $\beta$. Note also that high mass ratio systems and systems where the
Roche lobe is under-filled, will in general have smaller companions and higher 
effective temperatures. Of course all of these factors also
scale with the pulsar power and inversely with orbital separation $a$.

\begin{figure*}[ht!!]
\vskip 4.5truecm
\includegraphics{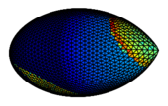}
\includegraphics{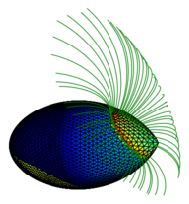}
\includegraphics{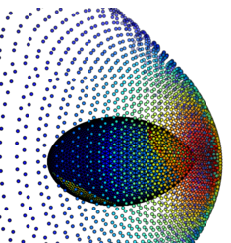}
\includegraphics{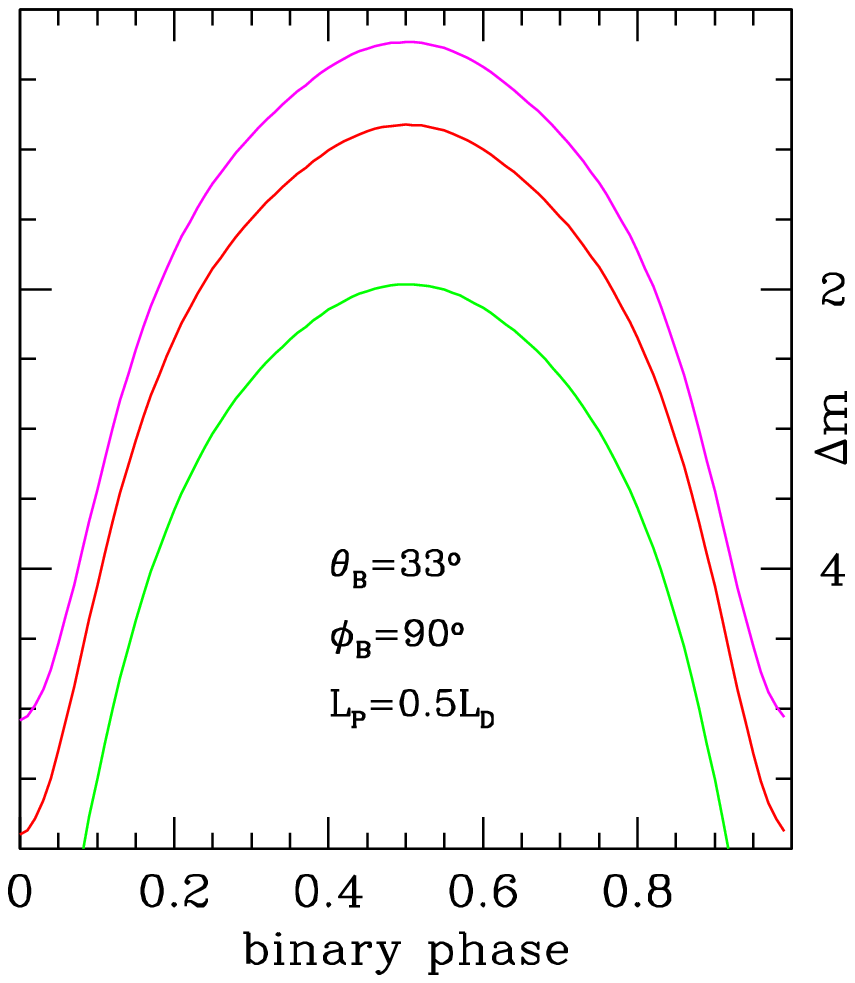}
\begin{center}
\caption{\label{Modelfigs} 
Left to right: a) IBS ($q=70$, $\beta=0.03$) and star ($f_c=0.99$), 
b) star and bounding field lines for front pole, 
c) heated star surface showing direct heating (faint blue), front cap with hot 
edge (red) and back partial cap (arc at bottom).
d) $g^\prime r^\prime i^\prime$ light curves for this model, viewed at $i=75^\circ$.
}
\end{center}
\vskip -0.5truecm
\end{figure*}

\section{Numerical Model}

	Although this simple model illustrates some basic scaling we do not use it
for direct data comparison. We instead assume an axially symmetric, equatorially concentrated
pulsar wind, typically
$$
f_P(r,\theta) = {\dot E(\theta)}/(4\pi r^2c) = 3I\Omega{\dot \Omega}{\rm sin}^2\theta/(8\pi r^2 c).
\eqno (9)
$$
but alternatively distributed as ${\rm sin}^4\theta$ \citep{tsl13}. We also
use the more detailed IBS shape of \citet{crw96}, 
$r(\theta) = d ~ {\rm sin}\theta_1/{\rm sin}(\theta + \theta_1)$ with $\theta$ the angle
subtended from the pulsar and
$$
\theta_1=\left [ {15\over 2} \left ( \left [
1+{4\over 5}\beta (1-\theta {\rm cot}\theta)\right ]^{1/2} -1 \right ) \right ]^{1/2}
\eqno (10)
$$
that from the companion star. For finite $f_V$ the IBS is swept back along the orbit.
See RS16 for details. Recently \citet{wet17} have presented a similar discussion of this
basic IBS geometry.

	The orientation of the companion magnetic dipole axis is specified by
$\theta_B$, $\phi_B$ (with 0,\,0 toward the pulsar). Since the pulsar heating may
be a significant driver of the dynamo-generated field, we allow the field origin to
be offset from the companion center. Here we only consider offsets along the line of
centers by $\lambda_B$, where  $\lambda_B=+1.0$ places the field origin at the 
companion `nose' closest to the $L_1$ point. If we displace the field, we can also
assume that the companion momentum flux, whether from a stellar wind or the dipole $B$ itself,
has a similar offset. Of course in general an offset $B$ might be centered 
anywhere within the companion, and might well include higher order multipoles.
Dipole dominance becomes an increasingly good approximation as the standoff distance
increases $r_0 > r_\ast$.

	In the numerical model we consider both direct photon heating and ducted
particle heating. The direct heating is assumed to be from 
the SED-dominating observed $\gamma$ ray flux $f_\gamma$, so that
at a point on the companion surface a distance $r$ from the pulsar with surface
normal inclination $\xi$ the direct heating flux is
$$
f_{\rm D} = L_{\rm D} {\rm cos}\,\xi/(4\pi r^2) =  f_\gamma (d/r)^2 {\rm cos}\,\xi.
\eqno (11)
$$
One caveat is that the observed $f_\gamma$ is measured on the Earth line-of-sight while the 
heating $\gamma$-rays are directed near the orbital plane. In \citet{rw10} and \citet{phgg16}
$\gamma$-ray beaming calculations were presented that in principle let one correct
the observed flux to the heating flux. The beaming models are less certain for the 
MSP treated here and we do not attempt such correction, but note
that for binaries observed at high inclination, the (presumably spin-aligned) MSP
likely directs more flux toward its companion than we observe at Earth.

	For the particle heating, we compute the IBS geometry and find the dipole field
lines that are tangent to this complex, possibly asymmetric, surface. These 
divide the IBS particles into those inside the curve of tangent intersection, 
which may couple to the `forward' pole from those outside the line which can
in principle reach the opposite `back side' magnetic pole. In fact, for each patch of
the IBS surface we determine the threading field line's footpoint on the polar cap
and deposit the appropriate fraction of the pulsar wind power on the star. $L_P$ is
the normalization for the total particle power (with a ${\rm sin}^2\theta$ distribution)
that reaches the companion surface; this can be compared with ${\dot E}$.
In general, dipole field line divergence ensures that the edges
of the cap collect more MSP spindown power per unit area and are hence hotter, 
although the details depend on the IBS geometry. Thus the generic geometry is an
edge-brightened cap offset from the companion nose (below the $L_1$ point).
Some IBS regions connect to the the pole on the `night' side of the star. 
The central field lines of this pole extend down-stream, away from the pulsar and
so do not intersect the IBS. The result is a partial ring, and the back pole is in general
much more weakly heated. Although this back-side heating is more sensitive to
the details of the IBS geometry and the topology of the magnetotail (so our
simple approximate solution is more likely to miss the detailed behavior), this backside
illumination often appears to improve light curve fits and serves, in some cases to
dominate the nighttime flux. This is interesting as it provides a non-hydrodynamic
mechanism to transport heat to the night side of the star. 

	This IBS-B model follows only the geometrical effects, the gross energetics of
the surface heating and the thermal re-radiation of the multi-temperature atmosphere.
Both the $\gamma$-rays and the $e^+/e^-$ are penetrating so the deep heating
approximation is good. We have ignored here the additional X-ray heating from the
IBS (or any Compton component), as it seems energetically sub-dominant (Table 1). We also
do not follow the detailed particle propagation or response of the companion field
to the pulsar wind, beyond a simple cut-off at the IBS location. Thus many details
which might be captured by a relativistic MHD numerical simulation are missing.
However our intent is to allow comparison with observational data and so the present 
simplification which allows a full model to be computed in seconds on a modern
workstation, is essential to allow model fitting and the exploration of parameter space.
The prime IBS-B feature is that it {\it collects} pulsar spindown power, focusing
to the companion, and does so in a way that can have strong asymmetries controlled
by the magnetic field geometry. This gives it the potential to explain puzzling high temperatures
and light curve asymmetries of black widow heating patterns.

\begin{figure*}[t!!]
\vskip 8.99truecm
\includegraphics{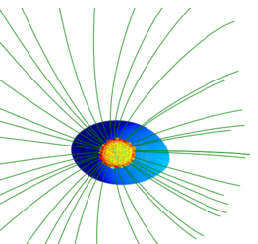}
\includegraphics{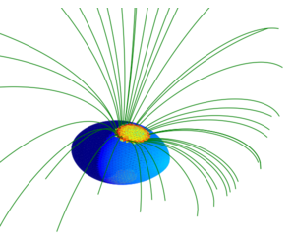}
\includegraphics{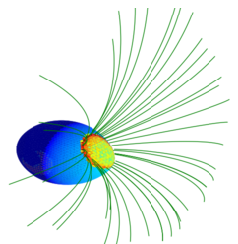}
\includegraphics{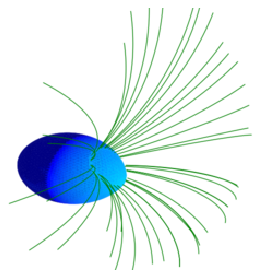}
\includegraphics{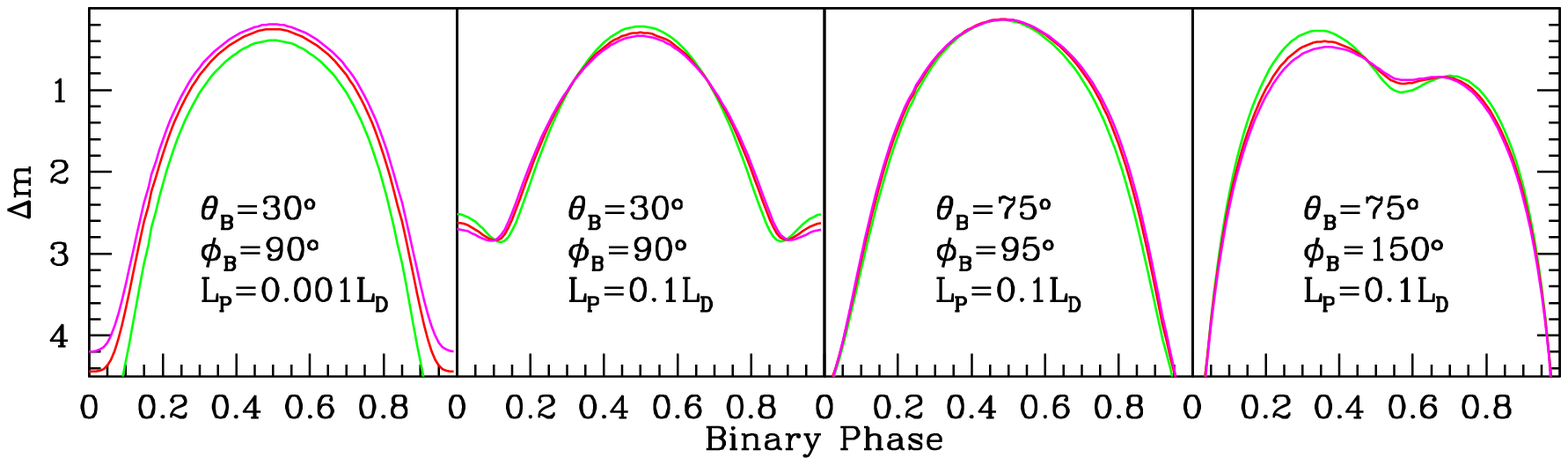}
\begin{center}
\caption{\label{Modelfigs} 
Example IBS-B geometries for $f_c=0.9$, $q=70$, $\beta=0.1$. Left to right a) Direct heating dominated
b) particle heating, small offset pole
c) large offset nearly aligned pole,
d) large offset, off-axis pole.
The bottom row shows the corresponding 
$g^\prime r^\prime i^\prime$ light curves, viewed at $i=75^\circ$. 
Note the bluer light curve colors from IBS heating starting with model b), the slight
color-dependent phase shift in model c) and the large peak distortion in model d).
}
\end{center}
\vskip -0.8truecm
\end{figure*}

Figure 2 shows the components of this model starting with the IBS,
then the dipole ducting field, the heated surface and the resultant light curves. Figure 3
shows an initial model with little particle heating compared with some sample magnetic geometries
and their resulting light curves.

\begin{deluxetable*}{lrrrrrrrrrr}[ht!!]
\tablecaption{\label{ObsPars} Pulsar Parameters}
\tablehead{
\colhead{Pulsar}&\colhead{${\dot E}_{34}^a$}&\colhead{P$_B$}&\colhead{$x_1/c$}&\colhead{DM}&\colhead{$\mu_T$} 
&\colhead{${\rm d_Y/d_N}^b$}&\colhead{$A_{V_Y}/A_{V_N}/A_{V_{\rm max}}$}&\colhead{$f_{\gamma,-11}$}&\colhead{$f_{X,-13}$}&\colhead{$K_{\rm obs}$}
\cr   
                &${\rm erg/s}$              &h               &lt-s    &${\rm cm^{-3}pc}$ &mas/y  
&kpc/kpc &  & ${\rm erg/cm^3/s}$ & ${\rm erg/cm^3/s}$ &km/s
}
\startdata
J1301+0833& 6.7&6.53&0.078& 13.2& 26.9&1.23/0.67&0.053/0.053/0.053&1.1& 0.3 &259\\
J1959+2048&16.0&9.17&0.089& 29.1& 30.4&1.73/2.50&0.806/1.054/1.364&1.7& 0.55&324\\
J2215+5135& 7.4&4.14&0.468& 69.2& ---&2.77/3.01&0.403/0.434/0.744&1.2& 0.7 &353\\
\enddata
\tablenotetext{a}{Before Shklovskii correction}
\tablenotetext{b}{Dispersion measure distance estimates from Y=YWM17 model, N=NE2001 model.}
\tablenotetext{c}{$K_{\rm obs}$ uncorrected sinusoid amplitude of companion optical radial velocity, see text.}
\end{deluxetable*}
\bigskip

\section{Light Curve Fits}

	The basic IBS structure is set by the dimensionless parameters 
$\beta$ and $f_v$, the magnetic field geometry by $\theta_B$ and $\phi_B$.
In practice the optical light curves are rather insensitive to $f_v$, so we
fix this parameter at a large value, leaving the bow shock symmetric. As noted in RS16
X-ray light curves are much more sensitive to $f_v$-induced asymmetry.
We also have the option of offsetting the field and wind centers from the
companion center by $\lambda_B$. In these computations we always apply the direct
heating of the observed $\gamma$-ray flux, as determined by fluxes from
the 2nd {\it Fermi} Pulsar catalog (or the 4th {\it Fermi} source catalog,
when pulse fluxes have not been published). This corresponds to
a direct heating power
$$
L_D=4\pi f_\gamma d^2 = 1.2 \times 10^{33}{\rm erg\, s^{-1}} f_{\gamma,-11} d_{\rm kpc}^2
\eqno (12)
$$
for typical parameters. This is often only a few percent of the 
characteristic spin-down power.
We chose here examples where there are published multi-color optical light curves
and radial velocity curves. Table 1 lists observed properties for our 
sample pulsars. 

As it happens, one critical parameter in the modeling is
the source distance. Early modeling work was largely based on the ELC
code, where only the shapes of the light curves in the individual
filters were considered. This provides immunity to zero point errors at the
cost of most color information and, with arbitrary normalization, is sensitive
to the source size only through shape (i.e. Roche lobe and tidal distortion)
effects. In our ICARUS-based model, the observed fluxes are used, making the
source size and distance important to the model fitting.
For these MSP the primary distance estimate is
from dispersion measure models. The NE2001 model \citep{ne2001} has become 
a standard reference, but the more recent \citet[hereafter YMW17]{ymw17} model
provides an updated comparison. The model uncertainties are hard to determine, but
a 20\% uncertainty is generally quoted. These MSP also have substantial proper 
motions and the distance is an important factor in determining Shklovskii effect
decrease to the spindown power
$$
{\dot E_{\rm Shk}}=4\pi^2 I v^2 /(c\,d\,P^2)
=9.6 \times 10^{33} I_{45} \mu_{10}^2 d_{\rm kpc} P_{\rm ms}^{-2}
\eqno (13)
$$
with the proper motion in units of 10\,mas/y.

	A final factor important to the optical modeling is the intervening extinction.
All of our sample objects are in the north, so we can use the 3-D
dust maps of \citet{gsf15} to infer $A_V$ at the source distance, and a maximum
$A_V$ along the pulsar line of sight.

\begin{figure}[t!!]
\vskip 8.9truecm
\includegraphics{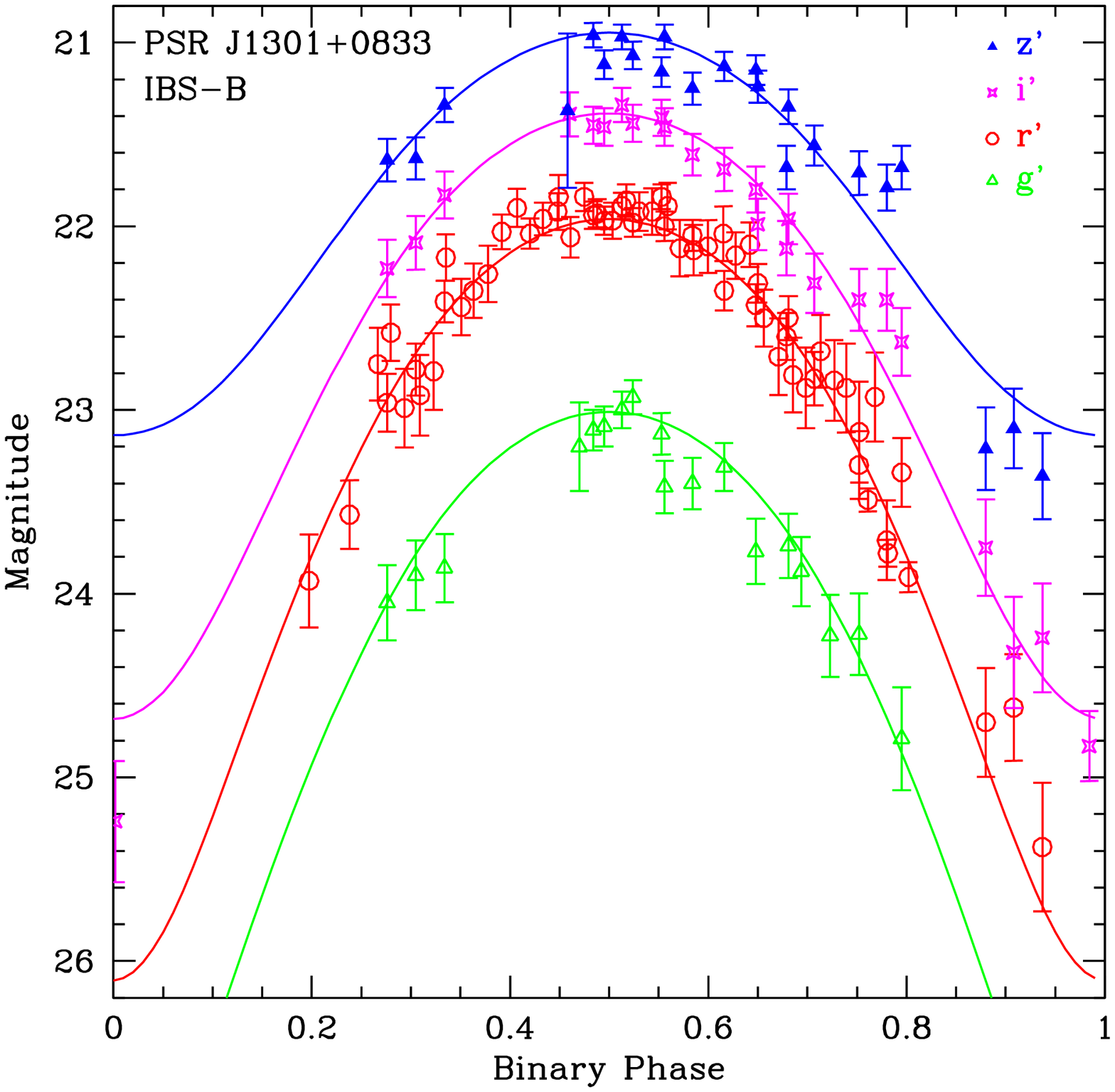}
\includegraphics{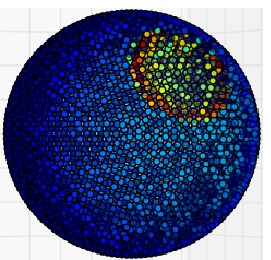}
\begin{center}
\caption{\label{J1301} 
IBS-B model fit to PSR J1301+0833. Inset shows the heating pattern with direct illumination
and the heated ring of the inner pole ($\theta_B=25^\circ, \, \lambda_B=-0.5$). 
}
\end{center}
\vskip -0.5truecm
\end{figure}

\bigskip
\subsection{Application to PSR J1301+0833}

This $P=1.84$\,ms BW has a $P_b=6.53$\,hr orbit with a $\sim 0.03 M_\odot$ companion.
We fit the 131 $g^\prime r^\prime i^\prime z^\prime$ photometric points
generated from a combination of imaging at MDM \citep{lht14}, Keck and Gemini archival
data and integration of Keck LRIS spectra (calibrated to a stable
neighboring star on the slit) described in RGFZ16.
The dispersion measure $DM= 13.2\,{\rm cm^{-3} pc}$ implies a distance 1.2\, kpc (YMW17)
or 0.67\, kpc (NE2001). As pointed out by \citet[RGFZ16]{rgfz16} the combination
of faint magnitude and modest $\sim 4500$\,K effective temperature requires a large 
$\sim 6$\,kpc distance for direct full-surface heating of a Roche-lobe filling star. However,
the pulsar has a substantial proper motion. The inferred space velocity and
Shklovskii-corrected ${\dot E}=6.7 I_{45} (1-0.31 d_{\rm kpc}) \times 10^{34}{\rm erg\, s^{-1}}$
are only reasonable for $\sim$kpc distances. So while the unconstrained direct model has 
a plausible  $\chi^2=184$ ($\chi^2/DoF=1.46$), the distance is $\sim 4\times$
larger than compatible with the proper motion (and DM).

	Thus to match the observed faint magnitude at the $\sim$ kpc distance the visible 
companion surface size must be substantially smaller than the volume 
equivalent Roche lobe radius $R_{\rm RL}$, {\it i.e.} fill factor $f_c = R_c/R_{\rm RL} << 1$.
For direct heating, at $d=1.2$\,kpc we need $f_c=0.137$ ($\chi^2=207$); 
at $d=0.67$\,kpc the size is $f_c=0.076$ ($\chi^2=216$). These are smaller than 
the fill factor of a cold $e^-$ degeneracy supported object of the companion mass, and 
so direct heating cannot provide a good solution unless $d>1.5$\,kpc.

	A magnetically ducted model provides a viable alternative, illuminating a small
cap of a larger star for a fixed $d=1.2$\,kpc.  For example, with a dipole field directed  
$\sim 25^\circ$ from the line of centers, offset $\lambda_B = -0.5$, we find a 
plausible $f_c=0.19$ model with $\chi^2=195$. Only a small fraction of the spin-down
power is required for this heating, with $\beta L_p = 1.0 \times 10^{30} {\rm erg\, s^{-1}}$. 
The light curve errors are too large for a tight independent constraint on $\beta$.
The area of the open zone above the caps scales with $\beta$ so $L_p \sim 1/\beta$.
For this small $\theta_B$ the back pole field lines 
are far off axis and we do not illuminate this pole as it overproduces flux at binary minimum.
The existing photometry strongly constrains the heated cap size and temperature, 
but the errors are too large to pin down the cap shape and location.
The spectroscopic points, in particular, may have substantial systematic errors away from companion maximum. 
Improved photometry, especially in the near-IR will help understand this system, and a good
X-ray light curve may help to measure $\beta$ and $f_v$.

\begin{figure}[t!!]
\vskip 8.9truecm
\includegraphics{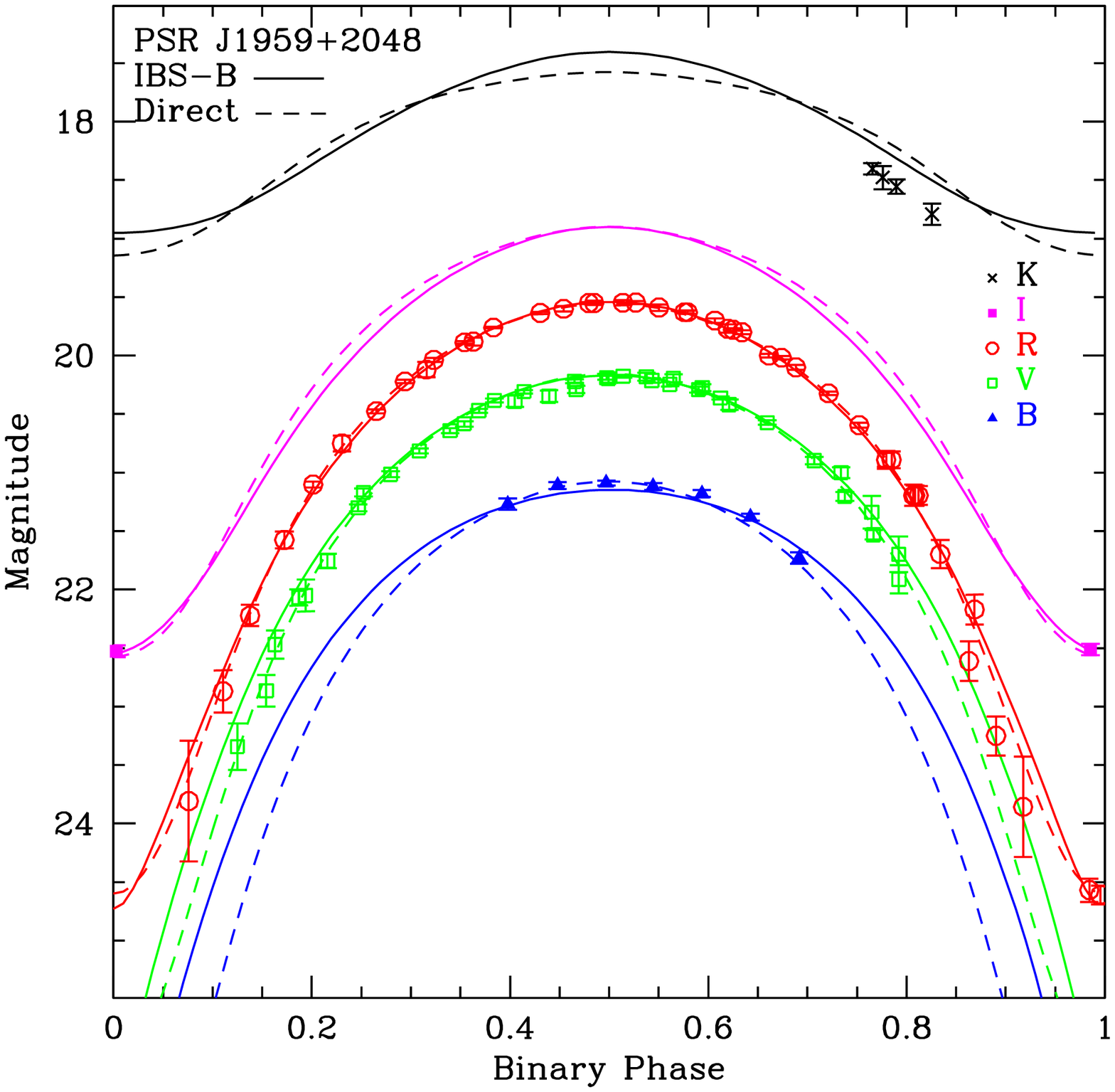}
\includegraphics{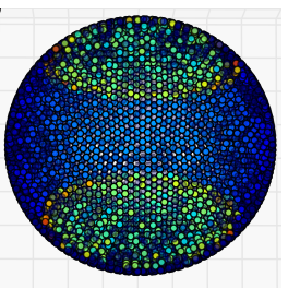}
\begin{center}
\caption{\label{J1959} 
IBS-B model fit to PSR J1959+2048 at $d=1.73\,$kpc (solid line) compared to the best 
unconstrained direct heating fits with $d=3.3$\,kpc (dashed line). Data from Reynold et al (2007).
Note the poor $K_S$ match. The inset show the heating pattern including the B-ducted heating.
}
\end{center}
\vskip -0.5truecm
\end{figure}

\subsection{Application to PSR J1959+2048}

	The original BW pulsar with $P=1.61$\,ms in a 9.2\,hr orbit is a prime target of these
studies, especially since \citet{vKBK11} measured a large radial velocity that, with existing
inclination estimates, imply a large neutron star mass. Here we use the photometry published in
\citet{ret07}; these 7\,B, 38\,V, 44\,R, 2\,I and 4\,K$_s$ points were kindly supplied by
M. Reynolds in tabular form. The I points and 2 R points at binary minimum are from {\it HST} WFPC
photometry.

	Again the source distance is a crucial uncertainty. The NE2001 and YMW17 distances
differ by 45\%. In fact the range may be even larger since \citet{vKBK11} estimated $d>2{\rm kpc}$
from spectral studies of the reddening, while \citet{arc92} estimated $d\sim 1.2\,{\rm kpc}$
based on spectroscopy of the H$\alpha$ bow shock. A direct heating fit produces a good match
to the multi-band light curves, with $\chi^2=256$. The error is dominated by the 4 $K_s$ points,
which might have a zero point error with respect to our model. A shift of
-0.3mag brings these into agreement, dropping $\chi^2$ to 164. However for this best fit the
distance is $3.3$\, kpc and the required extinction is larger than the maximum $A_V=1.36$ in 
this direction, which is only reached by $\sim$7\,kpc. Also, at this distance the Skhlovskii correction
decreases the spin-down power by 2/3 to $4.7\times 10^{34} I_{45}{\rm erg\, s^{-1}}$. The observed gamma-ray
flux corresponds to an isotropic luminosity of $L_\gamma = 2.2 \times 10^{34} {\rm erg\, s^{-1}}$,
a large fraction of the spin-down power, but even more disturbingly the sky-integrated luminosity
of the required direct heating flux is $9 \times 10^{34} {\rm erg\, s^{-1}}$, $4\times$ larger
than the observed flux and twice the inferred spindown power. Of course, if $I_{45} > 2$ and the
equatorial beaming increases the $\gamma$-ray flux above the Earth line of sight value by
4$\times$, then an efficient $\gamma$-ray pulsar can supply the required direct heating. Such a large 
$I_{45}$ would be of interest for the neutron star EoS and such high beaming efficiency would be
of interest for pulsar magnetosphere models. As an alternative the source might be closer.

	If we fix the distance at the YMW16 value of 1.73\,kpc and the corresponding $A_V=0.81$,
the fit is much worse, with $\chi^2=935$ (although the $K_s$ fluxes match!). However this model
requires a relatively small fill factor $f_c=0.31$ and an unacceptably small inclination 
$i=47^\circ$. Again the preference for smaller areas at larger temperatures implies 
concentrated heating and suggests that an IBS-B model can help.

We have indeed found an adequate ($\chi^2=295$) model for this distance/$A_V$ with the magnetic field 
pointing near the angular momentum axis, $\sim 90^\circ$ from the companion nose, and 
offset $\lambda_D= 0.3$ toward this nose.  This fit has a more reasonable $f_c=0.48$ and $i=76^\circ$, 
and, in addition to the $6.1\times 10^{33} {\rm erg\, s^{-1}}$ of direct heating it uses 
$3.8\times 10^{33} {\rm erg\, s^{-1}}$ of particle flux. This is $\sim (30I_{45})^{-1}$ 
of the available spindown power at this distance, suggesting that this fraction of the pulsar
at the open zone reaches the companion surface. The heated cap needs
to be rather large with $\beta \approx 0.005$ placing the IBS standoff at $\sim 7\%$ of the 
companion separation.
Our dipole approximation for the field geometry is suspect for such 
small standoff. Also, the X-ray orbital light curve of \citet{huet12} allows a larger $\beta$.
Figure 5 compares the direct (large d) and IBS-B models. As might be expected,
the largest differences lie in the infrared bands that detect more of the unilluminated surface.
Unfortunately the limited $K_s$ points here are not sufficient to control the model.
Happily van Kerkwijk and Breton have collected JHK observations of PSR J1959+2048 
(private communication); these will be very useful in choosing between 
the Direct-Free and IBS-B scenarios.

\subsection{Application to PSR J2215+5135}

	PSR J2215+5135 is a redback (RB) system, a $P=2.6$\,ms 
${\dot E}=7.4I_{45} \times 10^{34}{\rm erg\, s^{-1}}$ millisecond pulsar
in a $P_b=4.14$\,hr orbit with a $\sim 0.25 M_\odot$ companion. The LAT timing
model includes a $189\pm 23$ mas/y proper motion, but this is almost certainly spurious, 
due to the typical redback timing noise, as the space velocity would exceed 
100km/s at 110 pc. For this pulsar the two DM models give similar distances. 

	As in RS16 we fit the BVR light curves of \citet{sh14} 
(103 $B$, 55 $V$, and 113 $R$ magnitudes). The companion radial velocity
was measured in \citet{rgfk15}. As noted in RS16, direct fits are very poor
unless an (arbitrary) phase shift of $\Delta \phi \sim 0.01$ is imposed 
on the model. With such a phase shift the direct fit has a minimum $\chi^2=901$
($\chi^2=2268$ with no phase shift). The overall light curve shape and colors are
quite good, so this large $\chi^2$ is evidently the result of low level
stochastic flaring or underestimation of errors in the SH14 photometry (RS16).
The observed maximum is quite wide and to match this shape, the model prefers
$f_c \sim 0.9$ so that the ellipsoidal terms broaden the orbital light curve. 
In turn this requires a large distance $d_{\rm kpc}=5.6$. The best fit 
extinction is $A_V$=1.4, about twice the maximum in this direction. At this distance the
required heating luminosity $3.8 \times 10^{35} {\rm erg\,s^{-1}}$ is 
$5\times$ larger than the standard spindown power. 

	If we fix to the YMW17 distance and extinction, we find that the best fit 
direct heating model
adopts an unreasonably small inclination (and high irradiation power) to keep the
light curve maximum wide. In addition the fill factor $f_c=0.19$ is very small, inconsistent with
a main sequence companion.  The $\chi^2=2787$ is poor (and with no imposed phase shift
is $\chi^2=4208$), but larger inclination minima have $\chi^2$ many times worse. 

IBS-B models do better and can naturally produce the phase shift. At 2.77\,kpc we reach 
$\chi^2=3030$, but we find that the $T_N$ determined from the colors (and spectral type) 
at minimum coupled with the faint magnitude limit the emission area so that 
the fill factor is $\sim 0.41$, still small for a main sequence companion. It seems quite
robust that larger companions require $d_{\rm kpc}$ larger than the DM estimate.
If we free the distance, we find a $\chi^2=1318$ minimum
at 4.2\,kpc ($A_V=0.59$). The direct heating ($\gamma$-ray) power is 
$2.6\times 10^{34} {\rm erg\,s^{-1}}$ while the particle heating 
is only $2\times 10^{32} {\rm erg\,s^{-1}}$.
The $f_c=0.64$ implies a companion radius slight inflated from the main sequence expectation.
In RS16, IBS illumination models did produce lower $\chi^2$, but at $\sim 5$\,kpc distances
with large heating powers of $\sim 3 \times 10^{35} {\rm erg\,s^{-1}}$, even with
100\% efficient IBS reprocessing to companion-illuminating radiation. The
present solution, while statistically worse, seems physically preferable. The existing
X-ray light curve of PSR J2215+5135 is rather poor and the $\beta$ fit here is acceptable
(although finite $f_v$ is preferred by the X-ray data).

\begin{figure}[t!!]
\vskip 8.9truecm
\includegraphics{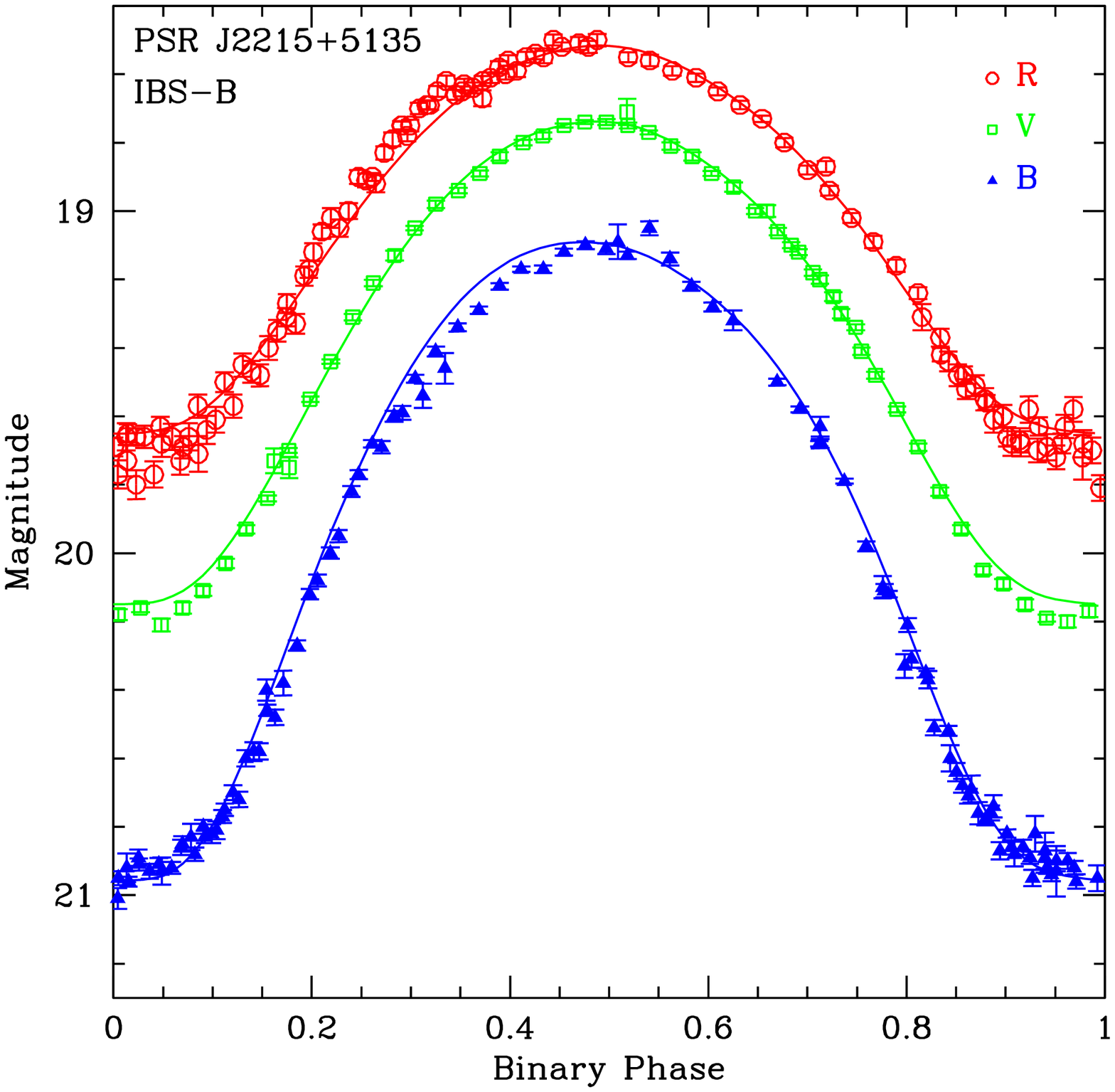}
\includegraphics{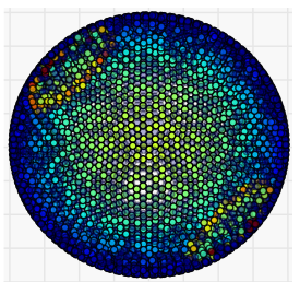}
\begin{center}
\caption{\label{J2215} 
IBS-B model fit to PSR J2215+5135. Inset shows the heating pattern with direct illumination
and two poles visible on the inner face ($\lambda_B=0.2$)
}
\end{center}
\vskip -0.5truecm
\end{figure}

\begin{deluxetable*}{lrrr|rrr|rrr}[ht!!]
\tablecaption{\label{FitPars} Model Fits$^a$}
\tablehead{
\colhead{Parameter}&\multicolumn{3}{c|}{J1301+0833}&\multicolumn{3}{c|}{J1959+2048}&\multicolumn{3}{c|}{J2215+5135}\cr
\colhead{}&\colhead{Direct-Free}&\colhead{Direct-Fixed}&\colhead{IBS-B}{ }
&\colhead{Direct-Free}&\colhead{Direct-Fixed}&\colhead{IBS-B}{ }
&\colhead{Direct-Free}&\colhead{Direct-Fixed}&\colhead{IBS-B}{ }
}
\startdata
$\chi^2$, DoF& 184/126&207/127&195/125&256/90&935/91&      295/89& 901$^d$/266&2787$^d$/267&1318/263\cr 
$f_{\rm cor}$&   1.095&   1.017&   1.026&    1.098&    1.035&   1.041&   1.077&   1.033&   1.052\cr 
$i$(deg) &   51(3)&   52(2)&   43(3)&    65(3)&  47.1(4)& 75.8(9)&   78(3)& 28.0(2)&   82(1)\cr 
$f_c$    & 0.68(8)&0.137(8)& 0.19(1)&  0.86(3)& 0.308(3)&0.476(4)&0.868(4)&0.187(1)& 0.64(1)\cr 
$T_N$(K) &2656(89)&2674(62)&2410(90)&2989(310)&2050(131)&2670(16)&7872(100)&4902(36)& 6120(9)\cr 
$L_D^b$  & 0.70(8)& 0.77(5)&     0.2&   9.0(9)&   6.1(1)&    0.61&37.6(48)& 29.6(4)&     2.6\cr 
$d$(kpc) &  4.7(2)&    1.23&    1.23&   3.3(2)&     1.73&    1.73&  5.9(1)&     2.8&  4.25(4)\cr 
$\beta^c$&        &        & 0.18[3]&         &         &0.005(1)&        &        & 0.47(1)\cr 
$\theta_B$(deg)&  &        &   25(4)&         &         & 89.9(8)&        &        & 87.5(9)\cr 
$\phi_B$(deg)&   &        &      90&         &         &       90&        &         &  135(2)\cr 
$L_P^b$  &        &      &5.3[4]$\times 10^{-4}$&  &   & 0.38(6)&        &        & 0.019(2)\cr 
$q$      &    45.3&    42.1&    42.4&     70.0&     66.0&    66.3&    6.42&    6.16&    6.27\cr 
$M_c(M_\odot)$&0.032(4)&{\it 0.026(2)}&0.041(7)&0.035(3)&{\it 0.059(1)}&0.026(1)&0.22(1)&{\it 1.83(4)}&0.20(1)\cr 
$M_P(M_\odot)$& 1.43(18)&{\it 1.10(9)}&1.75(30)  &2.46(18)&{\it 3.90(8)}&1.71(2)&1.40(5)&{\it 11.3(2)}&1.27(1)\cr 
\enddata
\tablenotetext{a}{() last digit(s) projected statistical errors from model fits. [] last digit single parameter errors.
Values without errors are fixed assumptions, except $f_{\rm cor}$ and $q$.}
\tablenotetext{b}{in $10^{34}{\rm erg\,s^{-1}}$. $L_D$ is fixed at $L_\gamma$ for IBS-B model. $L_P=0$ for direct models.}
\tablenotetext{c}{$f_v=\infty$ (symmetric shock) assumed for ISB-B models. Magnetic field offset $\lambda_B$=-0.5 (J1301), 0.3 (J1959), 0.2 (J2215).}
\tablenotetext{d}{J2215 Direct heating models include an arbitrary phase shift in the model 
light curve. $\chi^2$ without such shift are much larger.}
\end{deluxetable*}

\section{Models and Masses}
	
   For our example fits we have chosen systems with published optical radial velocity 
amplitudes, so we have additional kinematic constraints on the models and can 
use the photometric fit parameters to probe the system sizes and masses. It is important
to remember that the observed radial velocity amplitude $K_{\rm obs}$ (Table 1) 
is not the companion center-of-mass radial velocity. This needed quantity
is $K_{\rm obs} f_{\rm cor}$, with $f_{\rm cor}$ an illumination-dependent correction. We find
values ranging from $1.01 < f_{\rm cor} < 1.16$, where the largest values are for 
heating concentrated to the companion nose near the $L_1$ point.  Even for
a given illumination model $f_{\rm cor}$ depends weakly on the effective wavelength of the
lines dominating the radial velocity measurement. Values appropriate to the model
fits and the spectral types identified in the radial velocity studies are listed in Table 2.

	With $f_{\rm cor}$ in hand we can determine the mass ratio as
$$
q=M_P/M_c= f_{\rm cor} K_{\rm obs} P_B/(2\pi x_1).
\eqno (14)
$$
The companion mass is then
$$
M_c=4 \pi^2 x_1^3 (1+q)^2/(G\, P_B^2 {\rm sin}^3 i)
\eqno (15)
$$
where $i$ is also determined from the model fits. If we know the nature of the
companion we can then predict the minimum fill factor since we can compare with
the volume equivalent Roche lobe radius
$$
R_L=0.46 (1+q) x_1 q^{-1/3}/{\rm sin}\, i.
\eqno (16)
$$
For redbacks like PSR J2215+5135, we expect a main sequence companion with 
$R\approx (M/M_\odot ) R_\odot$.
For black widows, we might assume that the companions are solar abundance planetary
objects with radii $R \approx 0.135 (M/10^{-3} M_\odot)^{-1/8} R_\odot$, but if they are cold
degeneracy pressure supported evolved stellar remnants, they would have an
unperturbed radius $R=0.0126  (2/\mu_e)^{5/3} (M/M_\odot)^{-1/3} R_\odot$, with
$\mu_e=2$ for a hydrogen-free composition. In practice these radii should be viewed
as lower limits to the companion size, since it is widely believed that radiative
and tidal heating can inflate the companion stars.

	Given $P_B$, $x_1$ and $K_{\rm obs}$, for each $i$ and $f_{\rm cor}$ there is a
solution for the binary component masses and for the companion Roche lobe size.
Thus we can compare the model-fit $f_c$ with the minimum expected value given the
companion type. We show this comparison in Figure 7, where the curves give the 
expected size for $f_{\rm cor} = 1.02,1.08,1.12$. For J2215 we show the main
sequence prediction, while for J1301 and J1959 we show both the solar composition
planetary size and the H-free degenerate object size. The H-free
degenerate curves for the Tiddaren system PSR J1311$-$3430 are also plotted, to
show that large inclinations and near Roche-lobe filling are required for this binary.
The points with error flags show the model fits
discussed above. The direct unconstrained fits (triangle points) are all near Roche-lobe filling
and correspond to plausible neutron star masses, so these simple models would be
attractive if it were not for the substantial distance and power problems discussed
above. Direct fits locked to the DM-estimated distances (error bars without points)
are all at small $f_c$ and small $i$. The fill-factor error bars are quite
small as the source size is fixed given the temperature and flux, for a given distance.
For J1959 and J2215 the fixed-distance inclinations $i$ are so small that
the required neutron star mass is unphysical. For J1301 the mass is low and the companion
size is below even that of the cold degenerate model. The IBS-B solutions described
here are plotted as circle points. Note that all have plausible radii -- J1301 and J1959
are larger than the cold degenerate size and J2215 is just above the main sequence 
size.  The J2215 asymmetry is naturally produced by the offset magnetic poles and the 
required particle heating is less than 1\% of the spin-down power.
While the fit in Table 2 is the best found consistent with the physical constraints 
it does require a distance somewhat larger than the $DM$-estimated value. With the
large number of J2215 fit parameters it is not surprising that other (poorer) local minima 
exist in the fit, so the model conclusions must be considered preliminary. For example
larger $f_c$ can be accommodated if $d_{\rm kpc}$ is further increased.

	At the bottom of Table 2 we give the mass ratios and masses for these models,
assuming the parameters given in Table 1. We have put the `Direct-Fixed d' mass values
in italics since, as noted above, these are not good physical solutions. Both the 
`Direct-Free' masses and the `IBS-B' masses are plausible. The J2215 `IBS-B'
mass estimate is fairly
low for a neutron star but the J1301 and J1959 masses suggest at least modest accretion.

\begin{figure}[t!!]
\vskip 8.9truecm
\includegraphics{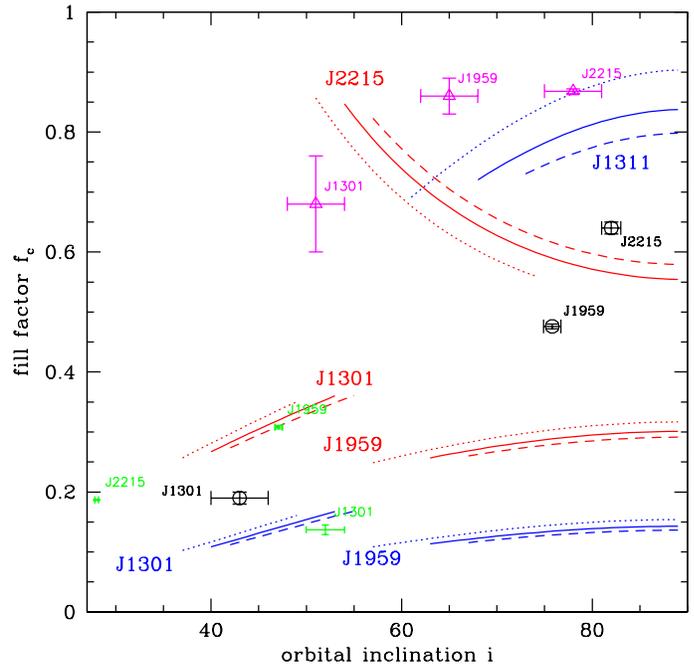}
\begin{center}
\caption{\label{ifcplane} 
Constraints and model fits in the inclination-fill factor plane. Curves show expected minimum
fill factors for the secondaries, with red for solar abundances and blue for H-free compositions. 
For J2215+5135, the curves show an unevolved MS secondary.
For PSR J1301+0833 and J1959+2048 we show solar abundance planetary radii (red) and H-free
degenerate radii (blue). Each curve is shown for $f_{\rm cor}=1.02$(dotted),
$f_{\rm cor}=1.08$(solid) and $f_{\rm cor}=1.12$(dashed) -- the curves range from $M_p=1.25 M_\odot$
to $M_p=2.5 M_\odot$. Fits for the unconstrained direct model (magenta triangle points) all
occupy a large fraction of the Roche lobe -- but imply improbably large distances. Direct 
heating model solutions at the DM-determined distances (green error flags) have 
unphysically large masses and small secondary radii.
IBS-B models (black circle points) are in agreement with, or inflated above, the expected
secondary size.
}
\end{center}
\vskip -0.5truecm
\end{figure}

\section{Conclusions}

	We have shown that the common picture of black widow heating, in which the companion is 
irradiated directly by high energy pulsar photons, often requires system distances 
substantially larger than allowed by other pulsar measurements and heating powers that
exceed the pulsar spindown luminosity. Note that this problem only comes to the fore when
one uses models that depend on the absolute band fluxes and is not obvious in other
(e.g. ELC) model fits.  This leads us to investigate a model in which
the direct heating by the pulsar $\gamma$-rays is supplemented by particle heating,
where the particles arise from pulsar wind reprocessing in an intrabinary shock and
then are ducted to the companion at its magnetic poles. This IBS-B ducting model 
has the potential to solve both the luminosity and distance problems as a larger
fraction of the spindown power is collected and concentrated to companion hot spots,
allowing a subset of the surface to dominate the observed radiation. This gives
smaller distances and more reasonable energetic demands. This is all very appealing
but one must ask how well it matches the data before trusting the fit parameters
and before embarking on detailed studies to assess the physical viability of this
(primarily) `geometry+energetics' model.

	We have applied this IBS-B model to several companion-evaporating pulsars.
Intriguingly, the best fits still occur for the simple direct heating models 
as long as the distance is unconstrained. In some cases it may simply be that the
DM-estimated distances are greatly in error and that the pulsar $\gamma$-ray radiation illuminates
the companion much more strongly than along the Earth line-of-sight. But in others such large
distances cannot be tolerated. If we require the distance to be consistent with DM
estimates, then our magnetic model can always provide a better fit than the direct heating
picture. In addition it can produce light curve asymmetries and required heating
powers consistent with those expected from the pulsar spindown. The pulsar masses
implied by the model fits are also much more reasonable if the close DM-determined
distances are adopted. But the fits are still imperfect and one may question 
whether this more complex magnetic model is warranted by the data. 

	Additional data can, of course, select between these models. Most important
are robust, independent distance determinations that set the size of the emitting area.
Unfortunately black widows and redbacks generally display too much timing noise to allow
timing parallaxes. However VLBI/GAIA parallaxes and, for systems like J1959, kinematic parallaxes
using the bow shock velocities and proper motions will be very valuable. In the X-ray
good measurement of caustics from the relativistic IBS particles can independently
constrain the $\beta$ and $f_v$ parameters, restricting the IBS-B model space. Improved
photometry of the companions, especially in the infrared where the weakly heated backside
can contribute will certainly help model fits. These factors will be particularly
helpful in modeling the important case of PSR J1959+2048. Finally very high quality phase-resolved
spectroscopy is sensitive to the distribution of the absorption lines over the visible face
of the companion (e.g. RGFK15). Sufficiently high quality companion observations must
reveal the details of the heating distribution.

	Our efforts have, so far, mainly increased the range of viable heating models.
We have highlighted problems with the standard direct heating assumption and have
produced a code that allows the IBS-B model to be compared with multiwavelength data.
This new modeling is important to black widow evolution and Equation of State (EoS) studies. 
For example, while the direct heating model implies a very large mass for PSR J1957+2048,
with dramatic EoS implication, the IBS-B fit value is more conventional.
On the other hand the IBS-B fit for PSR J1301+0833 show a larger, but uncertain,
neutron star mass.

As we apply this model to more black widow pulsars it should become clear what aspects of this
picture are robust and lead to improved understanding of the pulsar wind energy deposition.
Since the picture seems viable a more detailed analysis of the coupling between the pulsar
wind and the companion magnetosphere is needed; the examples in this paper suggest 
typical efficiencies of $\sim 1$\%.
The appeal of a good understanding of the wind's interaction with the companion is large, since
this provides a bolometric monitor of pulsar outflow many times closer than the X-ray PWN
termination shocks. This modeling also makes it clear that we need to get a clean understanding
of companion heating before we can make high confidence assertions about the dense matter
EoS.

\bigskip
\bigskip

We thank Paul Callanan and Mark Reynolds for allowing us to re-fit their photometry
for PSR J1959+2048, Alex Filippenko and his colleagues for their continuing interest in
the optical properties of black widow binaries, Hongjun An for discussions about
IBS shock physics, Rene Breton for advice on the ICARUS code and the anonymous referee
whose requests for clarification improved the paper. This work was supported in part
by NASA grant NNX17AL86G.

\end{document}